\documentclass[12pt,preprint]{aastex}

\shorttitle{Methanol Masers in G34.43+00.24 MM3}
\shortauthors{Yanagida et al.}

\begin{document}


\title{ALMA Observations of the IRDC Clump G34.43+00.24 MM3: 278 GHz Class I Methanol Masers}


\author{Takahiro Yanagida\altaffilmark{1}, Takeshi Sakai\altaffilmark{1}, Tomoya Hirota\altaffilmark{2,3}, Nami Sakai\altaffilmark{4}, Jonathan B. Foster\altaffilmark{5}, Patricio Sanhueza\altaffilmark{6,2}, James M. Jackson\altaffilmark{6}, 
Kenji Furuya\altaffilmark{7}, Yuri Aikawa\altaffilmark{8}, and Satoshi Yamamoto\altaffilmark{4}}

\altaffiltext{1}{Graduate School of Informatics and Engineering, The University of 
Electro-Communications, Chofu, Tokyo 182-8585, Japan.}
\altaffiltext{2}{National Astronomical Observatory of Japan, Osawa, Mitaka, Tokyo 181-8588, Japan.}
\altaffiltext{3}{Department of Astronomical Sciences, Graduate University for Advanced Studies, Mitaka, Tokyo 181-8588, Japan.}
\altaffiltext{4}{Department of Physics, Graduate School of Science, The University of Tokyo, Tokyo 113-0033, Japan.}
\altaffiltext{5}{Yale Center for Astronomy and Astrophysics, Yale University, New Haven, CT 06520, USA.}
\altaffiltext{6}{Institute for Astrophysical Research, Boston University, Boston, MA 02215, USA.}
\altaffiltext{7}{Leiden Observatory, Leiden University, P.O. Box 9513, 2300 RA Leiden, The Netherlands}
\altaffiltext{8}{Department of Earth and Planetary Sciences, Kobe University, Kobe 657-8501, Japan.}


\begin{abstract}
We have observed a molecular clump (MM3) associated with the infrared dark cloud G34.43+00.24 in the CH$_3$OH $J_K$=$9_{-1}$--$8_0$ $E$, $5_0$--$4_0$ $E$, and $5_{-1}$--$4_{-1}$ $E$ lines at sub-arcsecond resolution by using the Atacama Large Millimeter/submillimeter Array.
By comparing the CH$_3$OH $J_K$=$9_{-1}$--$8_0$ $E$ emission with the CH$_3$OH $5_0$--$4_0$ $E$ and $5_{-1}$--$4_{-1}$ $E$ emission, we have found that the CH$_3$OH $J_K$=$9_{-1}$--$8_0$ $E$ emission is masing.
We have clearly shown that the CH$_3$OH $J_K$=$9_{-1}$--$8_0$ masers arise from the post shocked gas in the interacting regions between the outflows and ambient dense gas.  Toward the strongest peak of the CH$_3$OH maser emission, SiO $J$=6--5 emission is very weak. This indicates that the CH$_3$OH maser emission traces relatively old shocks or weak shocks.

\end{abstract}


\keywords{ISM: clouds --- ISM: molecules --- stars: formation}



\section{INTRODUCTION}

Methanol masers are commonly found in star-forming regions, and are used as indicators of star formation activity (Menten 1991a).
Methanol masers can be classified into two classes: class I and II (Batrla et al. 1987; Menten 1991a). Class I methanol masers are recognized as shock excited masers, while class II methanol masers are pumped by infrared radiation.  Although the classification of each maser type was originally based on emitting regions of the masers, Slysh et al. (1994) and Kurtz et al. (2004) pointed out that the definition is better based on transitions.

It is well established that class II methanol masers are exclusive tracers of high-mass star formation (e.g., Menten 1991b; Minier et al. 2003; Ellingsen 2007; Sugiyama et al. 2008).
In contrast, class I methanol maser emission is found both in high-mass (e.g., Kurtz et al. 2004) and low-mass (e.g., Kalenskii et al. 2010) star forming regions.  Class I methanol masers are usually associated with molecular outflows (e.g. Kurtz et al. 2004; Chen et al. 2009).
Plambeck \& Menten (1990) and Cyganowski et al. (2009) found that class I methanol maser emission comes from the interface between outflows and ambient dense gas.
In addition, class I methanol masers are also found toward ultracompact H II regions (e.g. Fontani et al. 2010; Voronkov et al. 2010), where shocks caused by expansion of the ultracompact H II regions are responsible for excitation.
Thus, class I methanol masers are thought to be good tracers of various shocks. 

Although class I methanol masers are thought to trace shocked regions, their peak velocities are usually similar to that of the ambient dense gas (e.g. Menten 1991a).
A relationship between the maser emitting regions and the ambient dense gas is not very clear due to a lack of high spatial resolution observations.
In this Letter, we have studied the relationship between class I methanol maser emission and dense molecular gas toward G34.43+00.24 MM3 at sub-arcsecond resolution with Atacama Large Millimeter/submillimeter Array (ALMA).

G34.43+00.24 is an infrared dark cloud (IRDC; Garay et al. 2004; Rathborne et al. 2006; Sanhueza et al. 2010), which contains 9 molecular clumps (MM1-MM9; Rathborne et al. 2006). 
The distance to this object is derived to be 1.56 kpc from VLBI parallax observations (Kurayama et al. 2011), while the distance derived by using the Galactic rotation curve (e.g., Sakai et al. 2008; Sanhueza et al. 2012) and near-infrared extinction (Foster et al. 2012) ranges around 3-4 kpc.
This discrepancy is discussed in detail by Foster et al. (2012, 2014).
An ultracompact H II region is associated with MM2. Thus, this cloud is forming at least one, and probably several, high-mass stars.
G34.43+00.24 MM3 is the third most massive clump in this IRDC.
MM3 is less luminous in the infrared than MM1 and MM2, and is associated with a dark extinction region in the Spitzer 8 and 24 $\mu$m images.
There is no ultracompact H II region in MM3.

Sakai et al. (2013; Paper I) identified a very young hot core/outflow system in MM3 by using ALMA.
In addition, they found a possible evidence for past star formation activities, and suggested that this clump has already actively formed low-mass stars.
Recently, Foster et al. (2014) found another evidence of star formation activity in this clump from near-infrared $K$-band observations.
Thus, MM3 is a good target to investigate the relationship between star formation activity and class I methanol masers.
Although Chambers et al. (2009) reported the detection of 25 GHz class I methanol masers toward this source by using the Green Bank telescope, high-resolution observations of the class I methanol masers have not been reported so far.

We observed the CH$_3$OH $J_K$=$9_{-1}$--$8_0$ $E$ line toward G34.43+00.24 MM3 by using ALMA.
The $J_{-1}$--$(J-1)_{0}$ $E$ lines of CH$_3$OH are likely to be a class I maser line (e.g., Voronkov et al. 2012).
We compare the emission from the methanol masers with that of N$_2$H$^+$ and CS, in addition to shock-tracing molecules, such as SiO and thermal CH$_3$OH, in order to investigate the emitting regions of the class I methanol masers.
Although Zinchenko et al. (2012) reported the detection of the CH$_3$OH $J_K$=$9_{-1}$--$8_0$ $E$ line toward S255 with the Submillimeter Array, a detailed study of this transition has not been reported so far.
This Letter is the first ALMA observation reported for a millimeter class I methanol maser.

\section{OBSERVATIONS}

Our ALMA observations were carried out in August 2012.
The details of the observations are described in Sakai et al. (2013) and T. Sakai et al. (2014, in preparation).
In this Letter, we present the new data for CH$_3$OH $J_K$=$9_{-1}$--$8_0$ $E$ (278.30451 GHz; $E_u$$\sim$109 K; Band 7), $J_K$=$5_0$--$4_0$ $E$ (241.70022 GHz; $E_u$$\sim$48 K; Band 6), and $J_K$=$5_{-1}$--$4_{-1}$ $E$ (241.76722 GHz; $E_u$$\sim$40 K; Band 6).
Transition rest frequencies are taken from the CDMS catalog\footnote{http://www.astro.uni-koeln.de/cdms}.
The data were reduced by using the CASA software package.
The line images were obtained by CLEANing the dirty images after subtracting the continuum directly from the visibilities.
The velocity resolution of the data used in this Letter is 0.60 km s$^{-1}$.
The synthesized beam for the $J_K$=$9_{-1}$--$8_0$ $E$ transition is 0$^{\prime\prime}$.74$\times$0$^{\prime\prime}$.50 with a P.A. of 101.4$^{\circ}$, and for the $J_K$=$5_0$--$4_0$ $E$ and $J_K$=$5_{-1}$--$4_{-1}$ $E$ transitions is 0.79$^{\prime\prime}$$\times$0$^{\prime\prime}$.60 with a P.A. of 101$^{\circ}$.1.

\section{RESULTS}

\subsection{Integrated Intensity Images and Spectra}

Figure 1a shows the integrated intensity image of CH$_3$OH $J_K$ = $9_{-1}$--$8_0$ $E$.
The CH$_3$OH $J_K$ = $9_{-1}$--$8_0$ $E$ emission shows a clumpy structure, and there are several spatial peaks.
We identify 6 peaks (M1-M6) with intensity $>$ 100 mJy beam$^{-1}$.
Figure 2 presents the spectra of CH$_3$OH $J_K$ = $9_{-1}$--$8_0$ $E$ toward the 6 peaks.
The spectra consist of a relatively strong narrow component and a weak broad component.
The velocity width of the narrow component is less than 1 km s$^{-1}$.
This narrow linewidth is comparable to those of class I CH$_3$OH masers toward other objects (e.g., Kurtz et al. 2004; Chen et al. 2009).
The Gaussian fit parameters to the CH$_{3}$OH lines are listed in Table 1.

Figures 1b and 1c show the integrated intensity images of CH$_3$OH $J_K$ = $5_0$--$4_0$ $E$ and CH$_3$OH $J_K$=$5_{-1}$--$4_{-1}$ $E$, both of which are thermally excited (non-maser).
The spatial distributions of CH$_3$OH $J_K$ = $5_0$--$4_0$ $E$ and CH$_3$OH $J_K$ = $5_{-1}$--$4_{-1}$ $E$ look similar to that of CH$_3$OH $J_K$ = $9_{-1}$--$8_0$ $E$, except that the $5_0$--$4_0$ $E$ and $5_{-1}$--$4_{-1}$ $E$ emission is more extended.
In Figure 2, we present the spectra of all three methanol transitions.
The $5_0$--$4_0$ $E$ and $5_{-1}$--$4_{-1}$ $E$ intensities are weaker than that of $9_{-1}$--$8_0$ $E$ at the peak velocity of $9_{-1}$--$8_0$ $E$, although the upper state energies of $5_0$--$4_0$ $E$ and $5_{-1}$--$4_{-1}$ $E$ are lower than that of $9_{-1}$--$8_0$ $E$.

Assuming optically thin emission and local thermodynamic equilibrium excitation, we calculate the excitation temperature from the intensity ratios of ($5_{-1}$--$4_{-1}$ $E$)/($5_0$--$4_0$ $E$) and ($9_{-1}$--$8_0$ $E$)/($5_0$--$4_0$ $E$) at the peak velocity of the $9_{-1}$--$8_0$ $E$ emission.
The results are listed in Table 1.
The excitation temperatures derived from the ($5_{-1}$--$4_{-1}$ $E$)/($5_0$--$4_0$ $E$) ratio range from 19 K to 40 K, while the excitation temperatures derived from the ($9_{-1}$--$8_0$ $E$)/($5_0$-$4_0$ $E$) ratio are negative.
This indicates population inversion for the $J_K$ = $9_{-1}$--$8_0$ $E$ transition.
The compact distribution, the narrow velocity width, and the population inversion indicate that the observed CH$_3$OH $J_K$ = $9_{-1}$--$8_0$ $E$ emission is maser emission.

In contrast to the 6 peaks, toward the hot core, the peak intensity of $9_{-1}$--$8_0$ $E$ is weaker than that of $5_0$--$4_0$ $E$ and $5_{-1}$--$4_{-1}$ $E$ (Figures 1 and 2).
At this position, the excitation temperature derived from the ($9_{-1}$--$8_0$ $E$)/($5_0$-$4_0$ $E$) ratio is 79 K.
Thus, maser emission is either absent or very weak toward the hot core. 
In addition, the $9_{-1}$--$8_0$ intensities are weaker than or comparable to the $5_0$--$4_0$ $E$ and $5_{-1}$--$4_{-1}$ $E$ intensities for the broad velocity components.

Figure 1d shows the 1.3 mm continuum image superposed on the CH$_3$OH $J_K$ = $9_{-1}$--$8_0$ $E$ image.
As seen in this figure, it is clear that there are no continuum peaks toward M1-M6.
Figure 1e shows the CH$_3$OH $J_K$ = $9_{-1}$--$8_0$ $E$ color image overlaid with the CS contours.
The CS emission is likely tracing outflows (Paper I).
All of the CH$_3$OH $J_K$ = $9_{-1}$--$8_0$ $E$ peaks are indeed associated with the CS outflows, and some peaks are located on the edge of the CS outflows.

Figure 1f shows the color image of 2.2 $\mu$m $K$-band from Keck telescope (Paper I) overlaid with the CS contours.
Diffuse $K$-band emission due to the H$_2$ line at 2.12 $\mu$m (Foster et al. 2014) is seen near M2 and M3.
Voronkov et al. (2006) also reported class I CH$_3$OH masers associated with H$_2$ emission toward the high-mass source, IRAS 16547-4247.
In addition, Cyganowski et al. (2009) reported that methanol masers are seen on the edge of the extended green objects or "green fuzzies", which are diffuse sources with a 4.5 $\mu$m excess likely due to shock excited H$_2$ lines. 
M2 and M3 are probably similar to such situations. 
Although H$_2$ emission is not seen toward the northern outflow lobe, this is likely due to dust extinction as pointed out in Paper I.

In Figure 1f, we also find that none of the methanol masers peaks are associated with infrared point sources. 
Although infrared emission is seen toward M2, this is extended emission and not a point source.
In particular, no infrared emission is observed near M5, where the maser emission is the brightest.
We have checked that there is no peak of $Spitzer$ 24 $\mu$m toward M5.
Thus, the maser emission does not originate from regions very close to luminous young stellar objects.
This is consistent with a well known characteristic of class I methanol masers (Menten 1991a); they are collisionally excited rather than IR pumped.

\subsection{Velocity Structures}

To make the emitting regions of CH$_3$OH $J_K$=$9_{-1}$--$8_0$ $E$ clear, we present velocity channel maps of CH$_3$OH $J_K$=$9_{-1}$--$8_0$ $E$, CS $J$=5--4 and N$_2$H$^+$ $J$=3--2 (Figure 3), where the N$_2$H$^+$ data is presented in T. Sakai et al. (2014; in preparation). The N$_2$H$^+$ emission is known to trace cold gas (e.g., Tafalla et al. 2004).
At the velocity range of 58-60 km s$^{-1}$, we can see an anti-correlation between the N$_2$H$^+$ and CS emission. 
This indicates an interaction between the outflow and ambient dense gas, which is discussed in T. Sakai et al. (2014; in preparation).
In the panel at 58.07 km s$^{-1}$, we can see the strong peak of the CH$_3$OH $J_K$ = $9_{-1}$--$8_0$ $E$ emission toward the M1 position. This peak is clearly located on the edge of the N$_2$H$^+$ emission.
Thus, the maser emission comes from the interacting regions between the outflow and ambient cold dense gas.

In Figures 4a and 4b, we show the position-velocity (PV) maps of CS and SiO each overlaid on N$_2$H$^+$ and CH$_3$OH $J_K$ = $9_{-1}$--$8_0$ $E$ through the cut labeled \#1 in Figure 1e.
As reported in Paper I, the CS and SiO emission is broad ($\Delta V$$>$5 km s$^{-1}$) toward the shocked regions.
In these figures, the peak position of the methanol maser is almost spatially coincident with those of the CS and SiO emission, and it is offset from the N$_2$H$^+$ peak.
Thus, it is likely that the methanol maser emission comes from the post-shocked gas near the shock front.

In Figures 4a and 4b, the peak velocity of the methanol maser line is almost coincident with that of the N$_2$H$^+$ line.
In addition, the methanol maser emission is not strong in the broad high velocity components.
In Figure 4c, we present the PV map of the methanol thermal transition $5_0$--$4_0$ $E$. 
In this figure, the methanol thermal line is not strong in the broad components either, as compared with the SiO and CS lines.
The intensity ratio of the thermal CH$_3$OH and SiO emission is different by more than one order of magnitude between the narrow and broad components.
This may indicate that the abundance of CH$_3$OH is relatively low in the broad components.
Thus, the weakness of the maser emission in the broad components could originate from the less abundance CH$_3$OH.
It seems likely that CH$_3$OH is destroyed in high velocity shocks.
In fact, Garay et al. (2002) suggested that the CH$_3$OH molecule is destroyed for shock velocities above 10 km s$^{-1}$.

In Figure 3, strong methanol maser emission is seen toward M4 and M5 at the panel of 59.27 km s$^{-1}$.
However, we cannot see a clear interaction between the outflow and the cold dense gas toward M5.
In Figures 4d-f, we present the PV maps through the cut labeled \#2 in Figure 1e.
In Figures 4d-f, the methanol maser emission is found to be distributed along the N$_2$H$^+$ lane. Toward M4, the methanol maser peak is almost coincident with the N$_2$H$^+$ peak.
Although N$_2$H$^+$ emission is not seen toward M5, the peak velocity of the methanol maser emission agrees well with that of the N$_2$H$^+$ lane.
Thus, we suggest that the methanol maser emission at M5 is also related to the interaction between the cold ambient gas and some star formation activity.

\section{Discussion}

Here, we discuss the origin of the CH$_3$OH maser emission toward M5, the strongest maser in this source.
Toward M5, the SiO emission is very weak (Figure 4e), although the methanol thermal emission is rather strong (Figure 4f).
Such a trend is also seen toward M6 (Figures 4h-4i).
These characteristics are different from those observed in M1-M4.
Since the upper state energy and the critical density are comparable between the observed SiO and CH$_3$OH lines, this difference is likely to reflect a difference in their abundances.
Although the SiO emission is also weak toward the hot core, as compared with the methanol thermal emission (see also Paper I), no methanol masers are detected toward the hot core.
Thus, the situation is probably different between the hot core and M5.

The CH$_3$OH molecule is mainly formed on grain surface, while Si exists mainly in dust cores. 
Although Si is also thought to exist in ice mantles, the abundance would be much lower than that in dust cores (Jim\'{e}nez-Serra et al. 2008).
The radiation from YSOs can sublimate CH$_3$OH from the ice mantles at $>$90 K, but evaporation of dust cores requires much higher temperature ($>$1000 K).
Toward the hot core, the methanol emission traces such sublimated molecules.
The absence of methanol maser emission toward the hot core implies that all of the observed CH$_3$OH transitions are thermalized due to its high-density and high-temperature, as suggested by Cyganowski et al. (2011) in the hot core G19.01-0.03.

Toward M5, there are no detected 1.3 mm or infrared continuum sources.
Thus, the CH$_3$OH molecule in M5 cannot originate from radiative heating by embedded protostars.
Since M5 is located on the edge of the CS contours (see Figure 1),
the CH$_3$OH molecule in M5 seems to be released to the gas phase by shocks, as mentioned above.

One possibility is that the shock toward M5 is relatively old, and that SiO has already been depleted onto dust grains.
Since the sublimation temperature of SiO is much higher than that of CH$_3$OH ($\sim$90 K), SiO can deplete onto dust grains even in a hot ($>$$\sim$90 K) region.
As a result, gas-phase SiO could decrease faster than gas-phase CH$_3$OH in a post shocked region, unless the cooling timescale of post shocked materials is much shorter than the depletion timescale of gaseous molecules.
The depletion timescale of molecules depends on the gas density: $\sim$10$^9$/(n [cm$^{-3}$]) [yr] (Tielens, 2005).
Since the critical densities of the observed thermal CH$_3$OH lines are 10$^{6-7}$ cm$^{-3}$, the depletion timescale of CH$_3$OH could be 10$^{2-3}$ yr in the emitting region.
If the shock ended and the gas was cooled down below the sublimation temperature of CH$_3$OH about 10$^{2-3}$ years ago, CH$_3$OH would still be in the gas phase to some extent toward M5.
Cyganowski et al. (2012) also pointed out a similar possibility from the observations of class I methanol masers and SiO toward massive YSOs in G18.67+0.03.
However, it is unclear whether the maser emission can be maintained for a relatively long timescale after the shock ended. Model calculations may be necessary to assess this possibility.

Another possibility is that the methanol maser emission toward M5 traces relatively weak shocks.
Caselli et al. (1997) suggested that mantle evaporation occurs when the shock velocities are larger than 6 km s$^{-1}$, and grain-core destruction occurs when the shock velocities are larger than 19 km s$^{-1}$.
Thus, the CH$_3$OH/SiO abundance ratio depends on the strength of the shock, and may be high in a weak shock.
The high CH$_3$OH/SiO intensity ratio toward M5 may suggest that a weak shock occurs there, although we cannot find a clear velocity structure indicating an outflow near M5.

In either case, the driving source of the outflow is not clear.
According to the morphology of the CS emission, the Spitzer sources
might be the driving source. If we can investigate proper motions
of the maser spots by using VLBI (e.g. Matsumoto et al. 2014) and 
in the future obtain ALMA observations with longer baselines, we may find the driving source.
Such high resolution observations of methanol masers will be powerful tools to reveal a possible relationship between dynamical and chemical
evolution of cluster-forming regions.

\acknowledgments

This Letter makes use of the following ALMA data: ADS/JAO.ALMA\#2011.0.00656.S. 
ALMA is a partnership of ESO (representing its member states), NSF (USA) and NINS (Japan), together with NRC (Canada) and NSC and ASIAA (Taiwan), in cooperation with the Republic of Chile. 
The Joint ALMA Observatory is operated by ESO, AUI/NRAO, and NAOJ. We are grateful to the ALMA staffs. 
This study is supported by KAKENHI (21224002, 23740146, 24684011, 25400225 and 25108005).
JMJ acknowledges funding support from the US National Science Foundation via grant AST 1211844.
K.F. is supported by the Postdoctoral Fellowship for Research Abroad from the Japan Society
for the Promotion of Science (JSPS).

\clearpage



\begin{figure}
\epsscale{1.0}
\plotone{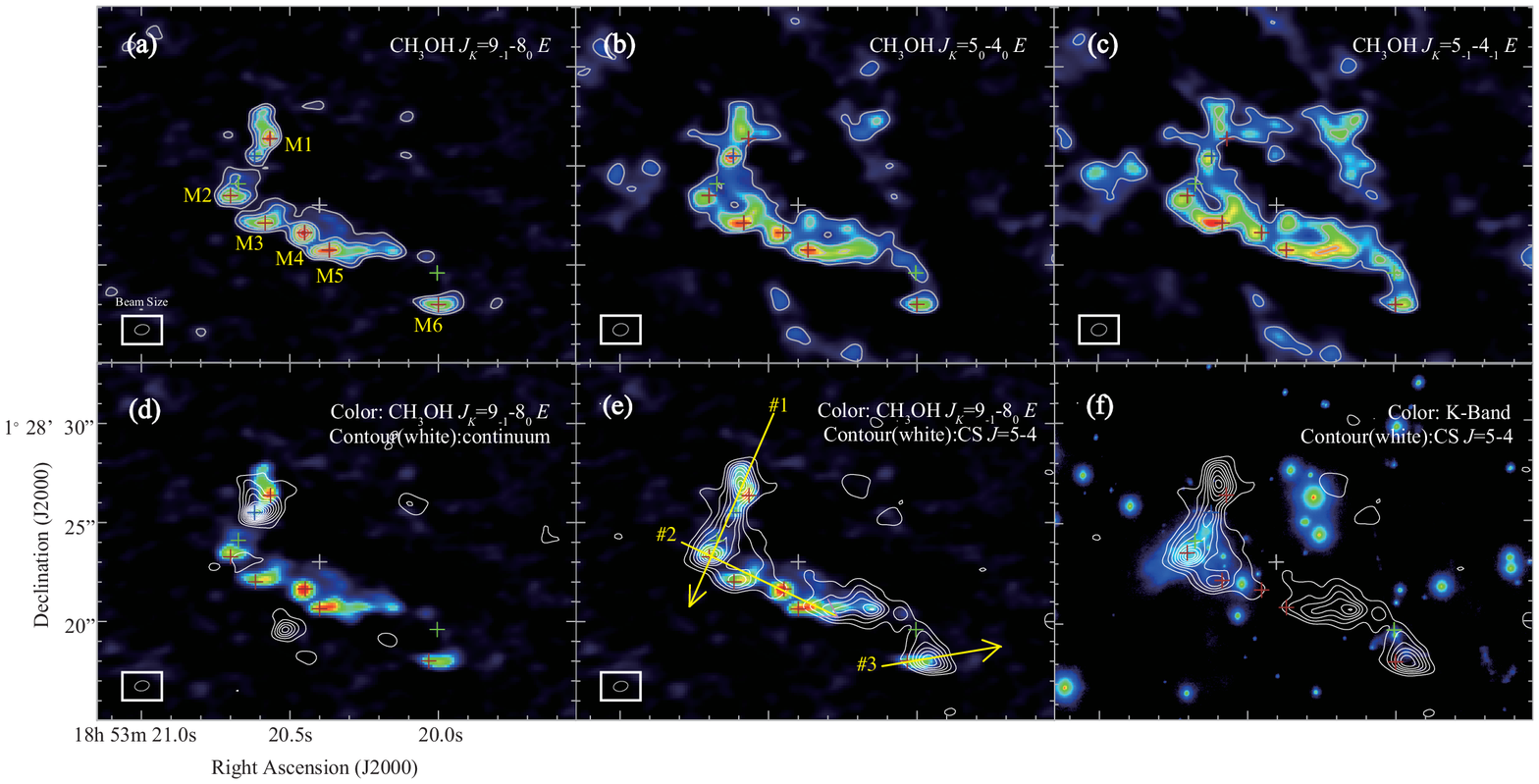}
\caption{Integrated intensity image and contours of (a) CH$_3$OH $J_K$ = $9_{-1}$--$8_0$ $E$, (b) CH$_3$OH $J_K$ = $5_0$--$4_0$ $E$, and (c) CH$_3$OH  $J_K$ = $5_{-1}$--$4_{-1}$ $E$.
(d) 1.3 mm dust continuum emission in contours (white contours) overlaid with the CH$_3$OH $J_K$ = $9_{-1}$--$8_0$ $E$ color image. (e) Integrated intensity of CS $J$=5--4 (white contours) superposed on the CH$_3$OH $J_K$ = $9_{-1}$--$8_0$ $E$ color image.  (f) Integrated intensity of CS $J$=5--4 (white contours) superposed on the $K$-band color image from Keck telescope.
The 1.3 mm continuum, CS, and $K$-band data are reported in Sakai et al. (2013). 
Contour levels start from 3$\sigma$ and increase in steps of 3$\sigma$ [(a) $3\sigma=255.0$ mJy beam$^{-1}$ km s$^{-1}$, (b) $3\sigma=240.0$ mJy beam$^{-1}$ km s$^{-1}$, (c) $3\sigma=450.0$ mJy beam$^{-1}$ km s$^{-1}$, (d) $3\sigma=1.2$ mJy beam$^{-1}$, (e) $3\sigma=480.0$ mJy beam$^{-1}$ km s$^{-1}$, (f) $3\sigma=480.0$ mJy beam$^{-1}$ km s$^{-1}$].  The white, blue, green and red cross marks represent the position of the phase center, the hot core, the Spitzer sources (Shepherd et al. 2007) and the CH$_3$OH $J_K$ = $9_{-1}$--$8_0$ $E$ peaks, respectively.  The solid lines indicated in (e) are for the position velocity diagram in Figure 4. 
\label{fig1}
}
\end{figure}

\clearpage


\begin{figure}
\epsscale{1.0}
\plotone{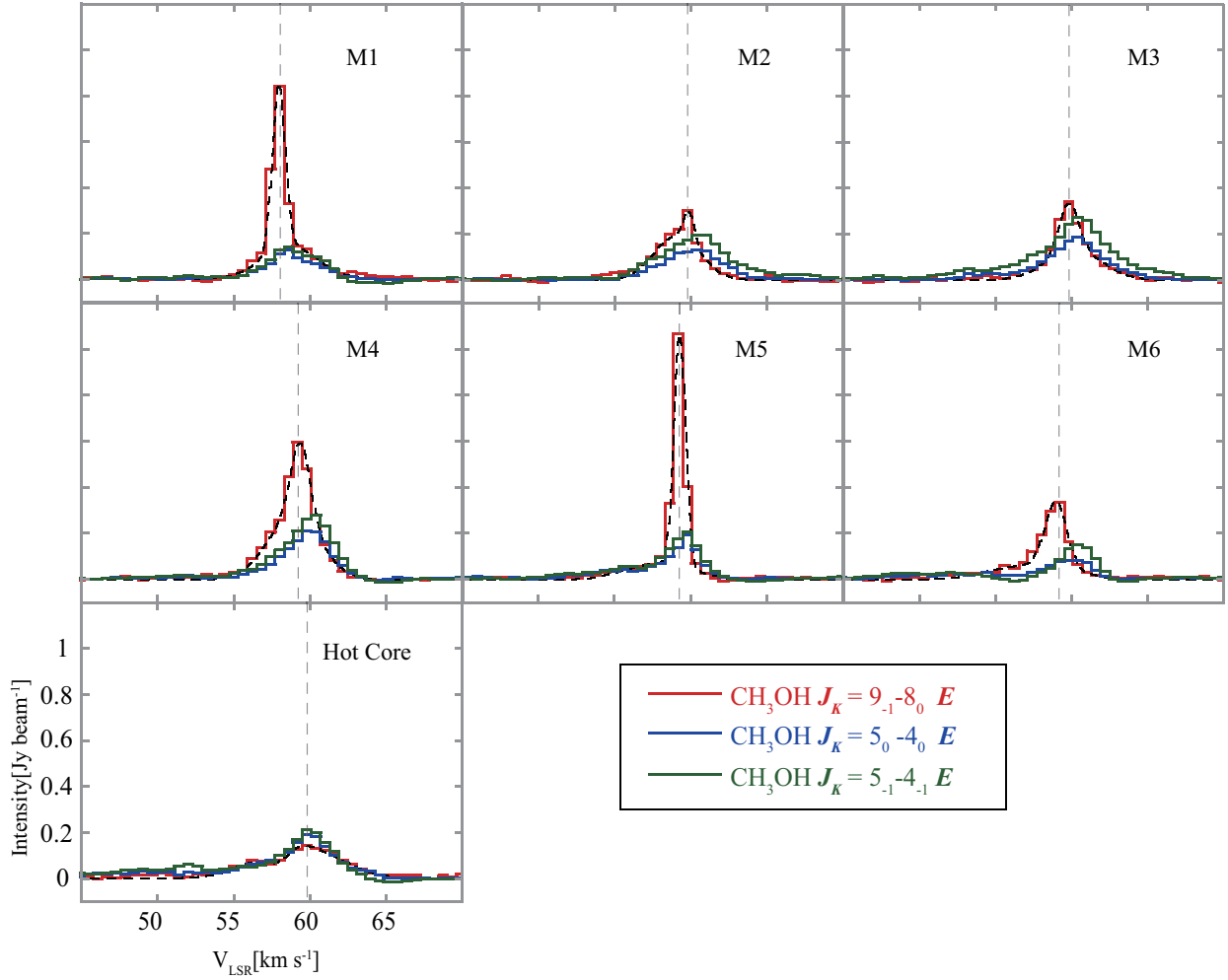}
\caption{
Spectra of CH$_3$OH $J_K$=$9_{-1}$--$8_0$ $E$, $J_K$=$5_0$--$4_0$ $E$, and $J_K$=$5_{-1}$--$4_{-1}$ $E$ toward the 6 peaks indicated in Figure 1a.
The black-dashed lines represent the results of the double Gaussian fitting to the $J_K$=$9_{-1}$--$8_0$ $E$ line.
The vertical grey dashed lines represent the peak velocity of the methanol maser emission.
An intensity of 1.0 Jy beam$^{-1}$ corresponds to 43 K.
\label{fig2}
}
\end{figure}

\clearpage

\begin{figure}
\epsscale{1.0}
\plotone{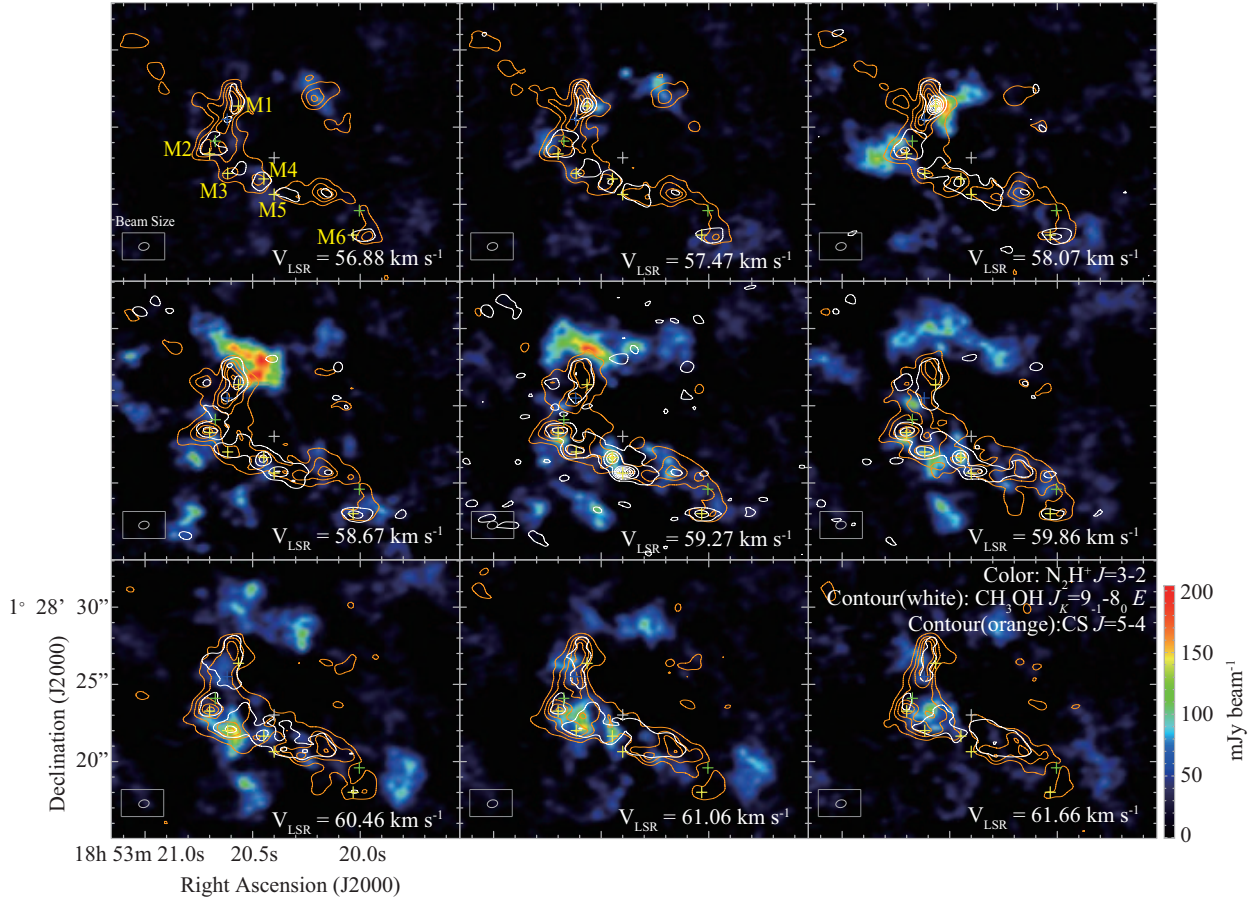}
\caption{
Velocity-channel maps of the N$_2$H$^+$ $J$=3--2 (color), CH$_3$OH $J_K$ = $9_{-1}$--$8_0$ $E$ (white contours), and CS $J$=5--4 (orange contours) within the velocity range from 56.88 km s$^{-1}$ to 61.66 km s$^{-1}$.
The lowest contour level and the contour step are (54.0 mJy beam$^{-1}$, 150.0 mJy beam$^{-1}$) and (42.0 mJy beam$^{-1}$, 70.0 mJy beam$^{-1}$) for CH$_3$OH $J_K$ = $9_{-1}$--$8_0$ $E$ and CS $J$=5--4, respectively. The cross marks are the same as Figure 1, but the methanol maser peaks are indicated in yellow.
\label{fig3}
}
\end{figure}


\clearpage

\begin{figure}
\epsscale{1.0}
\plotone{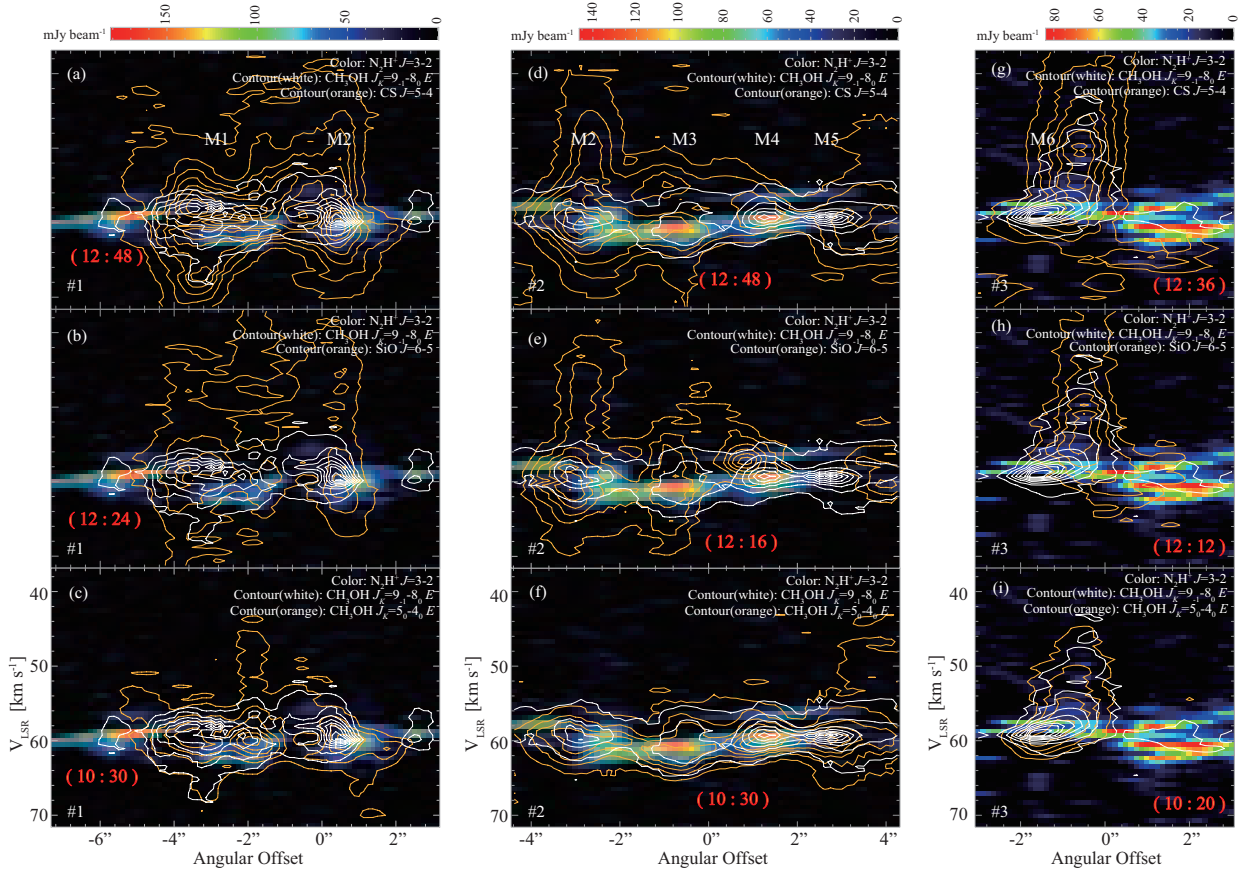}
\caption{
Position-velocity (PV) maps of CS $J$=5--4 (a), SiO $J$=6--5 (b), and CH$_3$OH $J_K$ = $5_0$--$4_0$ $E$ (c) each overlaid on N$_2$H$^+$ $J$=3--2 (color) and CH$_3$OH $J_K$=$9_{-1}$--$8_0$ $E$ (white contours) through the cut \#1 indicated in Figure 1e.
Figures d-f and g-i are the same as Figures a-c, but through the cut \#2 and \#3 indicated in Figure 1e, respectively
The lowest contour level and the contour step for CH$_3$OH $J_K$=$9_{-1}$--$8_0$ $E$ are (18 mJy beam$^{-1}$, 36 mJy beam$^{-1}$), (18 mJy beam$^{-1}$, 90 mJy beam$^{-1}$), and  (18 mJy beam$^{-1}$, 36 mJy beam$^{-1}$) for a-c, d-f, and g-h, respectively. 
The lowest contour level and the contour step for CS $J$=5--4, SiO $J$=6--5, and CH$_3$OH $J_K$ = $5_0$--$4_0$ $E$ are described in the each panel in units of mJy beam$^{-1}$. 
\label{fig4}
}
\end{figure}

\clearpage







\clearpage

\begin{deluxetable}{ccrrrrrrrrr}
\tabletypesize{\scriptsize}
\rotate
\tablecaption{Fit Parameters and Excitation Temperatures\label{tbl-1}}
\tablewidth{0pt}
\tablehead{
\colhead{No} & \colhead{R.A., Dec.(J2000)} &\multicolumn{2}{c}{$V_{\rm peak}$\tablenotemark{a}} &\multicolumn{2}{c}{$\Delta V$\tablenotemark{a}\tablenotemark{b}}& \multicolumn{3}{c}{Peak Intensity(CH$_3$OH)} & \multicolumn{2}{c}{$T_{\rm ex}$ ([$\cdot$]/($5_0$--$4_0$ $E$) )} \\
 &\colhead{18h53m[$\cdot$], 01d28m[$\cdot$]} & \colhead{Wide} & \colhead{Narrow} & \colhead{Wide} & \colhead{Narrow} & \colhead{$9_{-1}$--$8_0$ $E$} & \colhead{$5_0$--$4_0$ $E$}& \colhead{$5_{-1}$--$4_{-1}$ $E$} & \colhead{$9_{-1}$--$8_0$ $E$} & \colhead{$5_{-1}$--$4_{-1}$ $E$}  \\
 & & \footnotesize{[km s$^{-1}$]}  &\footnotesize{[km s$^{-1}$]} &\footnotesize{[km s$^{-1}$]} &\footnotesize{[km s$^{-1}$]} &  \footnotesize{[mJy beam$^{-1}$]} & \footnotesize{[mJy beam$^{-1}$]}& \footnotesize{[mJy beam$^{-1}$]} &  \footnotesize{[K]} &\footnotesize{[K]}\\
}
\startdata
M1 & 20.567s, 26.37s & 58.7 & 58.0 & 1.6 & 0.4 & 835 $\pm$18 & 104$\pm$3 & 122$\pm$3 & -39$_{-1}^{+1}$ & 39$_{-6}^{+9}$  \\
M2 & 20.700s, 23.25s & 59.1 & 59.9 & 1.3 & 0.3 & 300 $\pm$18 & 119$\pm$3 & 169$\pm$3 & -147$_{-26}^{+20}$  & 19$_{-1}^{+2}$  \\
M3 & 20.617s, 22.00s & 60.2 & 59.8 & 1.6 & 0.5 & 335 $\pm$18 & 166$\pm$3 & 236$\pm$3 & -315$_{-122}^{+72}$  & 19$_{-1}^{+1}$  \\
M4 & 20.450s, 21.63s & 58.9 & 59.4 & 1.5 & 0.5 & 591 $\pm$18 & 164$\pm$3 & 206$\pm$3 & -80$_{-4}^{+4}$  & 28$_{-2}^{+3}$  \\
M5 & 20.400s, 20.63s & 57.9 & 59.3 & 1.8 & 0.3 & 1063 $\pm$18 & 134$\pm$3 & 173$\pm$3 & -40$_{-1}^{+1}$  & 26$_{-2}^{+3}$  \\
M6 & 20.033s, 18.00s & 57.3 & 59.0 & 2.2 & 0.6 & 335 $\pm$18 & 72$\pm$3 & 83$\pm$3 & -60$_{-4}^{+4}$  & 39$_{-8}^{+15}$  \\
hot core & 20.617s, 25.50s & 59.2 & 57.8 & 2.3 & 0.7 & 140 $\pm$18 & 185$\pm$3 & 208$\pm$3 & 79$_{-11}^{+17}$  & 48$_{-6}^{+7}$  \\  
\enddata
\tablenotetext{a}{Results of the double Gaussian fitting of the 9$_{-1}$--8$_0$ $E$ line.}
\tablenotetext{b}{Corrected by the velocity resolution of 0.60 km s$^{-1}$ with the following equation: $\Delta V$=$\sqrt{\Delta V_{\rm fit}^2 - \Delta V_{\rm res}^2}$.}

\end{deluxetable}


\clearpage




\end{document}